\documentclass[review]{elsarticle}
\usepackage{graphicx}
\usepackage{lineno,hyperref}

\journal{Journal of \LaTeX\ Templates}

\begin{document}

\begin{frontmatter}

\title{Non-linear Thermoelastic Analysis of an Anisotropic Rectangular Plate}
\author{B. Das\fnref{myfootnote}}
\address{Department of Mathematics,\\ Ramakrishna Mission Vidyamandira,\\\it Belur Math, Howrah - 711202.}
\fntext[myfootnote]{~Corresponding author:
bappa.das1@gmail.com (B. Das).}

\author[mymainaddress]{S. Chakraborty and A. Lahiri}

\ead{csanjukta1977@gmail.com and lahiriabhijit2000@yahoo.com }

\address[mymainaddress]{Department of Mathematics,\\
\it Jadavpur University,Kolkata - 700032.}
\begin{abstract}
Three-dimensional thermoelastic analysis in presence of electro magnetic field is investigated of a rectangular plate. In the context of Green-Naghdi model-II, fractional order energy equation is adopted for a rotating anisotropic rectangular plate which is subjected to simply supported and isothermal on its four lateral edges. Normal mode analysis is adopted to the governing equations to formulate a Vector-matrix differential equation. The analytical closed form solution of the Vector-matrix differential equation is obtained for the physical parameters using eigen value approach methodology. Numerical results are represented graphically with a sinusoidal spatial variations of the stress applied on the top surface of the plate.
\end{abstract}

\begin{keyword}
\texttt{Eigenvalue}\sep  Fractional Order Energy Equation\sep Generalized Magnetothermoelasticity \sep Normal mode analysis and Vector-matrix
Differential Equation.
\end{keyword}

\end{frontmatter}


\section{Introduction}
Thermoelasticity is also widely spread term. It is used to refer various phenomena related to the interaction between deformation and heat conduction occurring in a body. Increasing attention has been devoted to this subject. Many researchers like Kawamura et al.\cite{R1} investigated the thermoelastic deformations of an orthotropic non-homogeneous rectangular plate. Reddy and Cheng \cite{R2} studied the three-dimensional deformations of functionally graded rectangular plate. Vel and Batra \cite{R3} also gave the three-dimensional exact solution for the vibration of functionally graded rectangular plate. Das and Lahiri \cite{R4},\cite{R5} illustrated the generalized thermoelastic solutions with or without electro magnetic field for functionally graded isotropic spherical cavity.

\paragraph{} In the engineering field, plates(rectangular or square, commonly used rectangular) are most widely used as structural materials. Such examples are Building Wall, Decks of Bridge, Airport and Highway Pavements etc. which are compatible to carry huge amounts of various loads.\\
To characterize the mechanical properties of the plate, it is important to classify the plates according to their thickness. One is of constant thickness and another is of variable thickness. Therefore, a total understanding of mechanical characterization of various engineering applications, it is also an widely receiving importance from material scientists and researchers. Zenkour \cite{R6} gave an exact solution for the bending of thin rectangular plate with uniform thickness. Although, a little number of cases have been studied for the solutions of variable thickness. Huang \cite{R7} studied the free vibration analysis of rectangular plate with variable thickness.

\paragraph{} Fractional calculus is a most important mathematical tools in the field of applied science and engineering. Many mathematicians like Liouville \cite{R8}, Riemann \cite{R9} etc. have been done vast research work on fractional calculus. Many researchers defined fractional calculus in different ways. Such as, Podlubny \cite{R10}, \cite{R11} suggested a solution of more than 300 years old problem of physical and geometrical interpretations of fractional differentiation and integrations. Carpinteri and Mainardi \cite{R12} solved the techniques of fractals and fractional calculus in Continuum Mechanics. The problems and solution procedure for differential equations of fractional order was proposed Kilbas and Trujillo \cite{R13}. Mainardi \cite{R14} analyzed the viscoelastic wave propagation using fraction calculus. Povstenko \cite{R15} also used time fractional derivatives to investigate thermal stresses in an infinite body with cylindrical hole.

\paragraph{} Many researchers have solved thermoelastic problems (classical, coupled, or generalized) with or without heat(body force) and/or with heat source by different ways such as: (i) state-space approach, this is essentially an expansion in a series in terms of the coefficient matrix of the field variables in ascending powers, which is the extensive application of Cayley-Hamilton theorem and (ii)eigenvalue approach, this method reduces the problem of a vector-matrix differential equation to an algebraic eigenvalue problem and the solutions for the resulting field equations are determined by solving these vector-matrix differential equations, which is the direct application of eigenvalues and the corresponding eigenvectors of the coefficient matrix. Liu and Zhong \cite{R16} discussed a three-dimensional thermoelastic analysis by state-space approach. Chakraborty, Das and Lahiri \cite{R17} have been solved many problems using eigen value approach methodology.

\paragraph{} In the present paper, we now introduce a three dimensional generalized magnetothermoelastic model with fractional order energy equation in
the context of Green-Naghdi model-II for a rotating anisotropic rectangular plate inclined with a constant angle with axis of rotation. For
mathematical computations, it is assumed that the plate is simply supported and isothermal on its four lateral edges.
Normal mode analysis is adopted to the governing equations to formulate a Vector-matrix differential equation. The analytical closed form solution of the
Vector-matrix differential equation is obtained for the physical parameters using eigen value approach methodology. Numerical computations are made
and represented graphically subjected to initial and boundary conditions.

\section{Theory of Generalized Magnetothermoelasticity with Fractional Order Energy Equation}
Assuming that a body having mass density $\rho$ is in unstressed and undeformed state at the constant reference temperature $\theta_0$ under the influence of thermal and mechanical stresses, the body undergoes a deformation $u_i$, (i=1,2,3) and respective temperature increment acquired ($\theta-\theta_0$), the absolute temperature $\theta$ is chosen such that $|\frac{\theta-\theta_0}{\theta_0}|<<1$. \\
In absence of internal heat source and body force, equations of motion are of the form Chakraborty, Das and Lahiri \cite{R17}
\begin{eqnarray}
\tau _{ij,j}  + {\bf{F}}_i = \rho[\ddot{u}_i+({\bf{\Omega}}\times({{\bf{\Omega}}\times{{\bf u}}}))_i+
({2{{\bf {\Omega}}\times\dot{{{\bf u}}}}})_i]
\end{eqnarray}
where $\tau _{ij}$'s are the stress tensor and ${\bf{F}}_i$'s are the Lorentz force components taken into account of the form
\begin{eqnarray}
{\bf{F}}_i=\mu_0({\bf{J}}\times \bf{H})_i
\end{eqnarray}
When the body exerted the electromagnetic field components, it follows the Maxwell's equations
\begin{eqnarray}
\textrm{curl} ~{\bf{h}} = {\bf{J}} + \varepsilon_0\frac{\partial {\bf{E}}}{\partial t},~~~
\textrm{curl} ~{\bf{E}} = - \mu_0\frac{\partial {\bf{h}}}{\partial t}\nonumber\\
\textrm{div} ~ {\bf{h}} = 0,~~~~~~~~{{\bf{E}} = -\mu_0(\dot{\bf{u}}\times \bf{H})}\nonumber\\
{\bf{B}} = \mu_0({\bf{H}}+ {\bf{h}}),~~~~~~~~{\bf{D}} = \varepsilon_0{\bf{E}}
\end{eqnarray}
where, $\mu_0, \varepsilon_0$ are the magnetic and the electric permeability, respectively.\\
It is considered that the body rotates uniformly with an angular velocity ${\bf{\Omega}}=(\Omega Cos\phi, \Omega Sin\phi, 0)$, where the angle of inclination is $\phi$ with the axis of rotation.Then the body maintains its rotation, so its equation of motion contains two additional terms, for the time varying motion, the centripetal acceleration denoted by the term (${\bf{\Omega}}$$\times$(${\bf{\Omega}}$$\times${\bf{u}}) and the term (2${\bf{\Omega}}$$\times$${\bf{\dot{u}}}$) is the coriolis acceleration.\\
The constitutive stress tensors are
\begin{eqnarray}
\tau _{ij} =c_{ijkl} e_{kl} -\beta _{ij} (\theta -\theta _{0}  +\tau _{1} \dot{\theta })\delta_{ij}
\end{eqnarray}
where, the coefficients $c_{ijkl}$ are the elastic module of the material, $\beta _{ij}$ are related to the relation $\beta _{ij}=c_{ijkl}\alpha _{kl}$ and $\tau _{1}$ is the constant defined as the mechanical relaxation time.\\

For any continuous function $f(\tau)$ on $b_1 \le \tau \le b_2$ and for any $\tau_1\epsilon(b_1,b_2)$, $Re ~\alpha >0$, the Riemann-Liouville definition of fractional order integral (Liuville \cite{R8} and Riemann \cite{R9}) is given by
\begin{eqnarray}
 \tau_1D_t^{-\alpha}f(t)=\tau_1I_t^{\alpha}f(t)=\frac{1}{\Gamma (\alpha)} \int _{\tau_1}^{t}{(t-\tau )^{\alpha-1}}{f(\tau)}d\tau ,  0 \le \alpha \le 1
\end{eqnarray}
where,$\tau_1D_t^{-\alpha}f(t)=\tau_1I_t^{\alpha}f(t)$ is the Riemann-Liouville integral operator. Substitution of $\tau_1=0$ in equation (6) gives the definition of Riemann fractional order integral and substitution of $\tau_1=-\infty$ in equation (6) gives the definition of Liouville fractional order integral.\\
The non Fourier energy equation using Riemann fractional order integral is given by
\begin{eqnarray}
 I_t^{\alpha-1}K_{ij}^*\nabla ^{2}\theta = \rho C_{E} \theta_{,tt}+ \theta _{0}\beta _{ ij}(u_{i,j})_{,tt}
\end{eqnarray}
where, $K_{ij}^*$ is the Green-Naghdi's parameter, $\beta _{ ij}$ is termed as the thermal modules. This energy equation is also valid for an anisotropic body and are termed as the Duhamel-Neumann equations with fractional order time derivative.\\

\section{Formulation and solution of the Problem}
We consider an anisotropic rectangular plate occupying the region R =\{$x_1,x_2,x_3$ : $0\le x_1\le h$; $0\le x_2\le c$; $0\le x_3\le d$\} as shown in Fig1. Assuming the rotational vector ${\bf \Omega}$ makes an angle $\phi$ with $Ox_2$ edge i.e., ${\bf \Omega}=(\Omega_1,\Omega_2,0)$, where, $\Omega_1=\Omega Cos\phi, \Omega_2=\Omega Sin\phi$. Due to the constant initial magnetic field ${\bf{H}}=(0,H_0,0)$, it experiences induced magnetic field ${\bf{h}}=(0,h,0)$ and similarly experiences an induced electric field ${\bf E}=(E_1,0,E_3)$ and corresponding current density ${\bf{J}}=(J_1,0,J_3)$ also considered the plate is simply supported and isothermal on all four edges. The top and bottom surface of the plate is subjected to a prescribed mechanical and thermal load.
\vspace{5em}\\
\textheight 22.0cm
\begin{minipage}{1.5\textwidth}
\vspace{.8cm}
\includegraphics[width=3.65in,height=1.0in ]{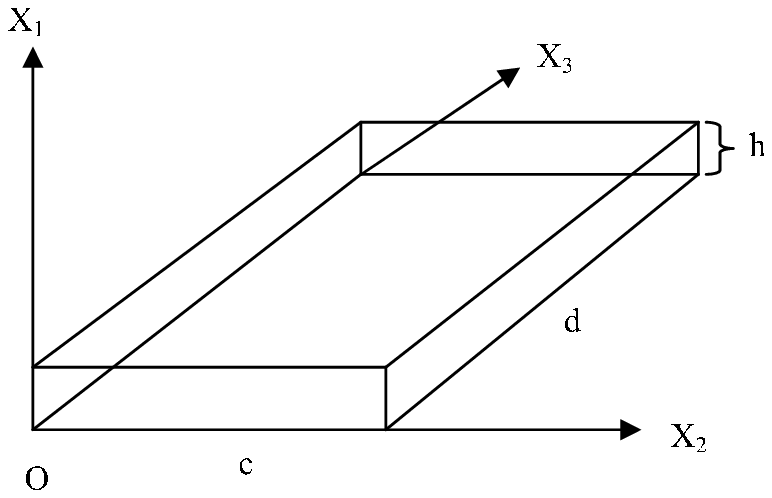}\\
Fig. 1 ~~Schematic representation of a rectangular plate\\
\end{minipage}\vspace*{.5cm}\\
From equation (3), we get the expressions as
\begin{eqnarray}
{\bf{h}}=-H_0(0,0,e) ; {\bf E}=\mu_0H_0(\dot{u}_3,0,\dot{u}_1) ; e=u_{j,j} ; {\bf{J}}=\{H_0(u_{1,x_1x_3}+u_{2,x_2x_3}+u_{3,x_3x_3})\nonumber\\-\mu_0H_0\varepsilon_0u_{3,tt},0,
-H_0(u_{1,x_1x_1}+u_{2,x_1x_2}+u_{3,x_1x_3})+\mu_0H_0\varepsilon_0u_{1,tt}\} ;\nonumber\\ {\bf{F}}=\{\mu_0H_0^2(u_{1,x_1x_1}+u_{2,x_1x_2}+u_{3,x_1x_3})-\mu_0^2H_0^2\varepsilon_0u_{1,tt},0,\nonumber\\
\mu_0H_0^2(u_{1,x_1x_3}+u_{2,x_2x_3}+u_{3,x_3x_3})-\mu_0^2H_0^2\varepsilon_0u_{3,tt}\}\nonumber\\
\end{eqnarray}
where, subscript variable denotes the differentiation.\\
The non-dimensional equations of motion and heat conduction equation will be obtained by using equation (7) in equations (1) and (6)
by the following non-dimensional variables
\begin{eqnarray}
{\rm (x}^{{\rm '}} _{{\rm i}} ,u^{{\rm '}} _{{\rm i}} )=\frac{1}{l}{\rm (} {\rm x}_{{\rm i}} ,\frac{\rho c_{1}^{2} }{\beta_{11} \theta _{0} } u_{{\rm i}} ), (t^{'},\tau _{i}^{'}) =\frac{c_{1} }{l}(t,\tau _{i}) ,  (\theta ^{'},\tau _{ij} ^{'}) =\frac{1}\theta _{0}({{\theta -\theta _{0} }} ,\frac{\tau _{ij} }{\beta_{11} }),\nonumber\\ e_{ij} ^{'} =\beta_{11} \theta _{0} e_{ij} ,\Omega_i  =\frac{c_{1} }{l} \Omega_i^{'}
\end{eqnarray}
where,  $l$ = $\frac{k*}{\rho c_{E} c_{1} } $, $c_{1} ^{2} =\frac{c_{11} }{\rho } $, $c_{1} $ represents the dilatation wave velocity, $\varepsilon_1$=$\frac{\mu_0H_0^2}{c_{11}}$, $\varepsilon_2$=$\frac{\mu_0H_0^2\varepsilon_0}{\rho}$, $\beta_2$=$\frac{\beta_{22}}{\beta_{11}}$,  $\beta_3$=$\frac{\beta_{33}}{\beta_{11}}$, $\varepsilon$=$\frac{\beta^2_{11}\theta_0}{c_{11}\rho c_E}$, $k_2$=$\frac{k^*_{22}}{k^*_{11}}$ and  $k_3$=$\frac{k^*_{33}}{k^*_{11}}$.

The physical variables occur in the equations of motion, energy and stress components can be decomposed in terms of normal mode forms which is given by
\begin{eqnarray}
\xi(x_i,t)=\xi^*(x_i,t)(x_{1} )e^{\omega t+i(ax_{2} +bx_{3} )}
\end{eqnarray}
where, $\xi(x_i,t)=[u_{i},e_{ij},\theta,\tau _{ij}]$,  $\xi^*=[u^*_{i},e^*_{ij},\theta^*,\tau^* _{ij}]$, $\omega$ is the angular frequency and $a,b$ are the wave numbers along $x_2$ and $x_3$ directions respectively.\\
Using equation (9), non-dimensional form of equations (1) and (6) can be written in the form of Vector-matrix differential equations as in Chakraborty, Das and Lahiri \cite{R17}
\begin{eqnarray}
D\underline{v}(x_1,t) = \underline{A}~\underline{v}(x_1,t); D\equiv \frac{d}{dx_{1} }
\end{eqnarray}
where, $\underline{v}$ =$\left[u_{1} {\rm \; \; \; }u_{2} {\rm \; \; \; }u_{3} {\rm \; \; \; }\theta {\rm \; \; \; }Du_{1}  {\rm \; \; \; \; }Du_{2}  {\rm \; \; \; }Du_{3}  {\rm \; \; }D\theta  {\rm \; }\right]^{T} $    and the coefficient matrix  $\underline{A}$ = $\left[{L_{ij} }\right]_{i,j=1,2}$

\noindent Where $L_{11}$ and $L_{12}$ are null and identity matrix of order 4x4 respectively and $L_{21}$ and $L_{22}$ are given in Appendix.

Similarly, equation (4) becomes
\begin{eqnarray}
\tau=hu+\Theta
\end{eqnarray}
where, $\tau$ =$\left[\tau_{ij}\right]^{T}_{i,j=1(1)3}$, $h$=$\left[{h_{ij} }\right]_{i,j=1(1)6}$,${u}$ =$\left[Du_{1}{\rm \; \; \; }Du_{2} {\rm \; \; \; }Du_{3} {\rm \; \; \; }u_{1}  {\rm \; \; \; \; }u_{2} {\rm \; \; \; }u_{3} {\rm \; \; }{\rm \; }\right]^{T} $,\\
 ${\Theta}$ =$\left[\theta{\rm \; \; \; }\beta_{2}\theta{\rm \; \; \; }\beta_{3}\theta{\rm \; \; \; }0{\rm \; \; \; \; }0 {\rm \; \; \; }0{\rm \; \; }{\rm \; }\right]^{T} $ and the values of $h_{ij}$ are given in Appendix.\\
The eigenvectors $X$=$\left[\delta _i {\rm \; \; \; }\lambda \delta _i\right]_{i=1(1)4}$ corresponding to the eigenvalues $\lambda $= $\lambda _{i}$, $(i=1(1)4)$ can be calculated from the characteristic equation
\begin{eqnarray}
\left|\left. \underline{A}-\lambda I\right|\right. =0
\end{eqnarray}\\

As in Chakraborty, Das and Lahiri \cite{R17}, the general solution of the equation (10) is
\begin{eqnarray}
{(u_i,\theta)}(x_1)=\sum _{j=1}^{8}A_{j} (\delta_{ij},\delta_4)\exp (\lambda _{j} x_{1} ) {\rm \; }  , x_{1} \geq0, \delta_{ij}=[\delta_i]_{\lambda=\lambda_j}
\end{eqnarray}
 The arbitrary constants $A_{j} $'s, $j$=1(1)8, are to be determined from the boundary conditions of the problem.

The expressions of the stress components are
$\tau=RA $, where, $R=[R_{ij}]_{i=1(1)6,j=1(1)8}$ and $A=[A_j]^T_{j=1(1)8}$,
where, $R_{ij}$, $\delta _{i}$ and the arbitrary parameters $A_j$ are given in Appendix.\\
The equation (7) also gives the expressions for magnetic field as well as the components of Lorentz force.\\

\section{Boundary Conditions}
For an anisotropic rectangular plate, it is considered that the plate is simply supported and isothermal on all four edges, the boundary conditions at edges are as follows-
\begin{eqnarray}
\tau_{22}=u_3=u_1=\theta=0    ~~~~at~~x_2=0~~and~~ c\nonumber\\
\tau_{33}=u_2=u_1=\theta=0    ~~~~at~~x_3=0~~and~~ d
\end{eqnarray}
The bottom and top surface of the plate experienced mechanical load as follows-\\
At $x_1=0$(bottom surface)
\begin{eqnarray}
\tau_{13}=X^-(x_2,x_3),~\tau_{12}=Y^-(x_2,x_3),\tau_{11}=Z^-(x_2,x_3), \theta=\theta^-(x_2,x_3)=0
\end{eqnarray}
At $x_1=h$(top surface)
\begin{eqnarray}
\tau_{13}=X^+(x_2,x_3),~\tau_{12}=Y^+(x_2,x_3),\tau_{11}=Z^+(x_2,x_3), \theta=\theta^+(x_2,x_3)=0
\end{eqnarray}
There is a sinusoidal mechanical load $(Z^+(x_2,x_3)=Z_0 Sin(\frac{\pi x_2}{c})Sin(\frac{\pi x_3}{d}), Z_0=-1 GPa)$ on the top surface of the rectangular plate, while other mechanical and thermal loads vanish on the top and bottom surfaces of the plate.

\section{Numerical Results and Discussions}
With a view of illustrating the problem, we now consider a numerical example for which computational results are presented. Since $\omega$ is complex, we take $\omega =\omega _{0} +i\varsigma $, for studying the effect of wave propagation, we use the following physical parameters in SI units given in the Table.\\
\noindent
\begin{center}
\noindent
$\underline{Table~~ of~~ physical~~ parameters}$
\noindent
\end{center}
\begin{center}
\begin{tabular}{|p{3.5in}|} \hline
$c_{11} $= 106.8 GPa ;$c_{22} $ = 99.00 GPa ;$c_{33} $= 54.57 GPa ;  $c_{12} $ = 27.10 GPa ;$c_{13} $= 9.68  GPa  ; $c_{14} $ = -0.03 GPa ;$c_{15} $= 0.28 GPa ;  $c_{16} $=  0.12 GPa ; $c_{21} $=27.10 GPa ;  $c_{23} $= 18.22 GPa  ;  $c_{24} $=1.49 GPa;  $c_{25} $= 0.13 GPa ; $c_{26} $ = -0.58 GPa ; $c_{32} $=18.22 GPa ; $c_{34} $=2.44 GPa  ;    $c_{35} $=  -1.69 GPa ; $c_{36} $= -0.75 GPa ; $c_{41} $=-0.03 GPa ;  $c_{42} $=1.49  GPa ; $c_{43} $=2.44 GPa  ; $c_{44} $ = 25.97 GPa ; $c_{45} $ = 1.98 GPa  ;  $c_{46} $= 0.43 GPa ; $c_{51} $=0.28 GPa ;$c_{52} $=0.13 GPa ; $c_{53} $=-1.69 GPa ; $c_{54} $=1.98 GPa ; $c_{55} $=25.05 GPa  ;  $c_{56} $=1.44 GPa ; $c_{61} $=0.12 GPa ;     $c_{62} $= - 0.58 GPa ; $c_{63} $=-0.75 GPa ; $c_{64} $=0.43 GPa ; $c_{65} $=1.44 GPa ; $c_{66} $= 37.82 GPa   ;   $\rho $= 2.727 ; $\varepsilon $ =0.334 ; $\tau _{1} $=0.1 ; $\tau _{2} $=0.1. \\ \hline
\end{tabular}
\end{center}
  \begin{center}
\includegraphics[width=4.8in,height=3.5in ]{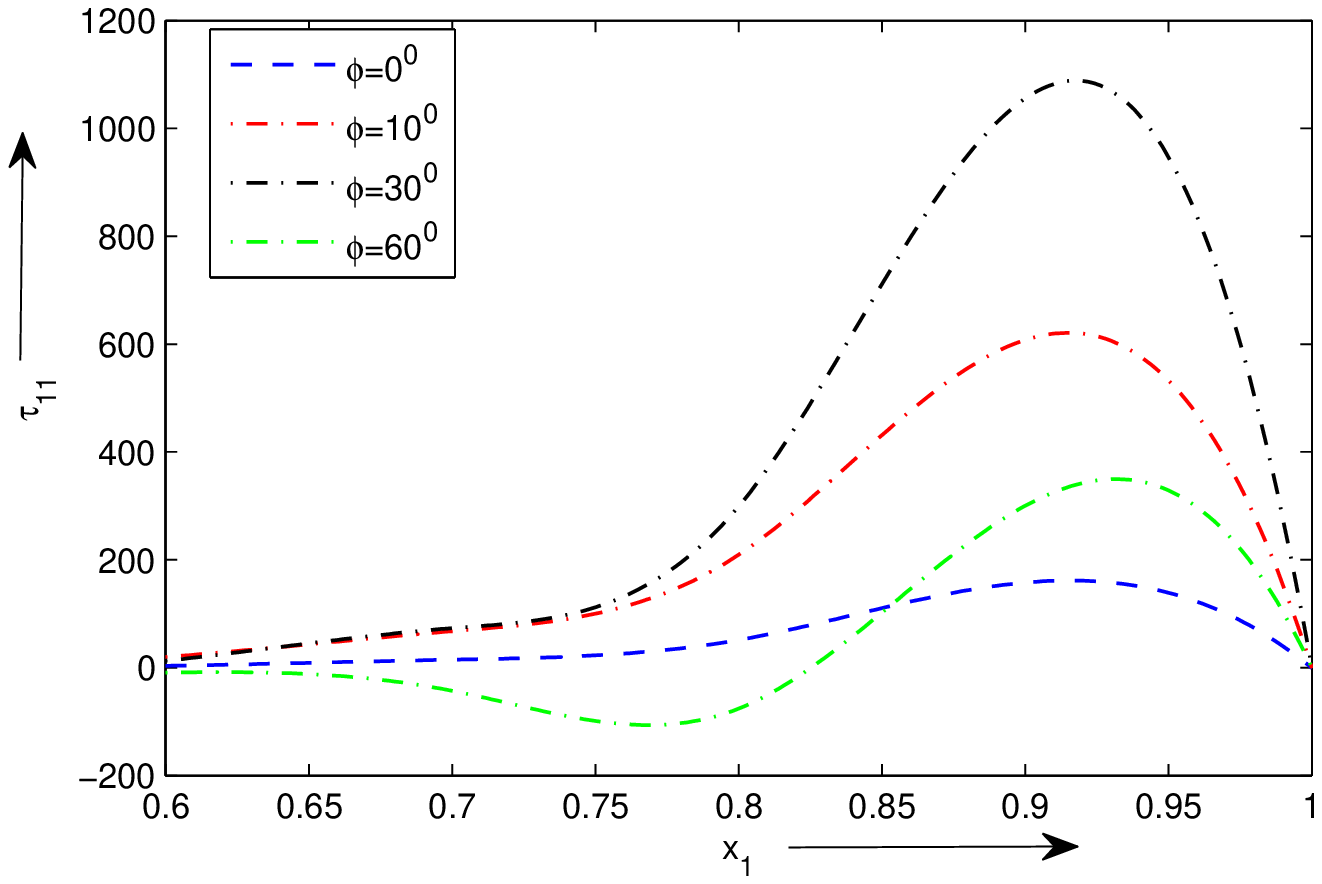}\\
Fig. 2 ~~Distribution of ($\tau_{11}$) vs. $x_1$ \\
\includegraphics[width=4.8in,height=3.5in ]{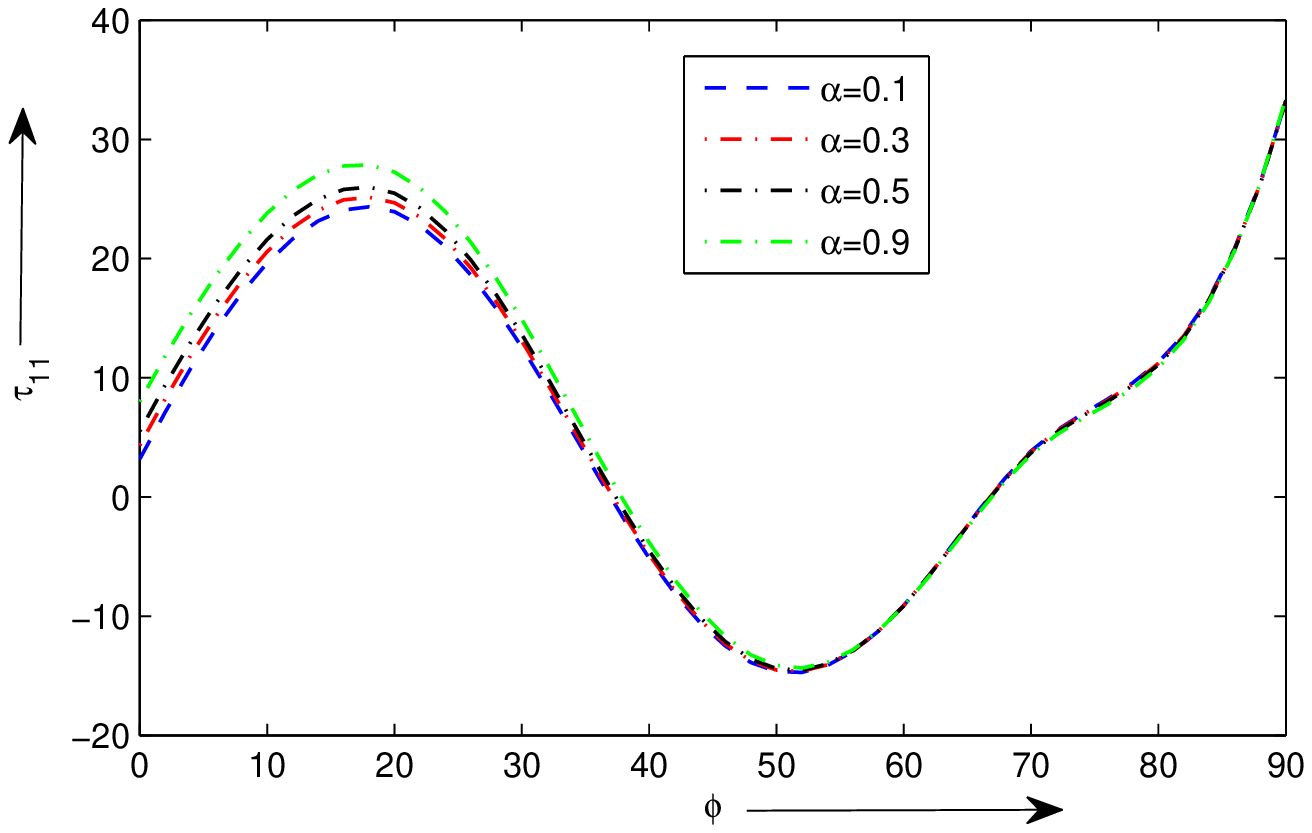}\\
Fig. 3~~Variation of ($\tau_{11}$) vs. $\phi$ \\
\end{center}

Figure 2 describes the distribution of $\tau_{11}$ versus space variable $x_1$ for fixed values of $\alpha=0.1$, time$(t)=0.5$ and four values of $\phi$, i.e.,$\phi=0^0, 10^0, 30^0, 60^0$. It is seen that numerical value of $\tau_{11}$ is minimum when rotational direction is along $x_2$ -axis i.e., $\phi=0^0$. Absolute value of $\tau_{11}$ gradually increases with the inclination angle $(\phi)$ and space variable $x_1$ and finally, the stress component $\tau_{11}$ becomes zero on the top of the plate. It is also seen that for $\phi=0^0, 10^0, 30^0$, stress($\tau_{11}$) is always extensive in nature for all space variable $x_1$ and it is compressive in nature for inclination angle$(\phi)=60^0$ within the region $0.7\leq x_1\leq 0.85$.  Figure 3 also describes variation of $\tau_{11}$ versus inclination angle($\phi$) for four fixed values of non-linear parameter($\alpha$)i.e.,$\alpha=0.1, 0.3, 0.5, 0.9$ also for fixed value of $x_1=0.6$ and time($t)=0.5$\emph{}. It is seen that nature of $\tau_{11}$ remains same for all values of $\alpha$ exceptionally, variation of $\alpha$ is prominent within $0^0 \le \phi \le 35^0$.\\

   \begin{center}
 \includegraphics[width=4.8in,height=3.5in ]{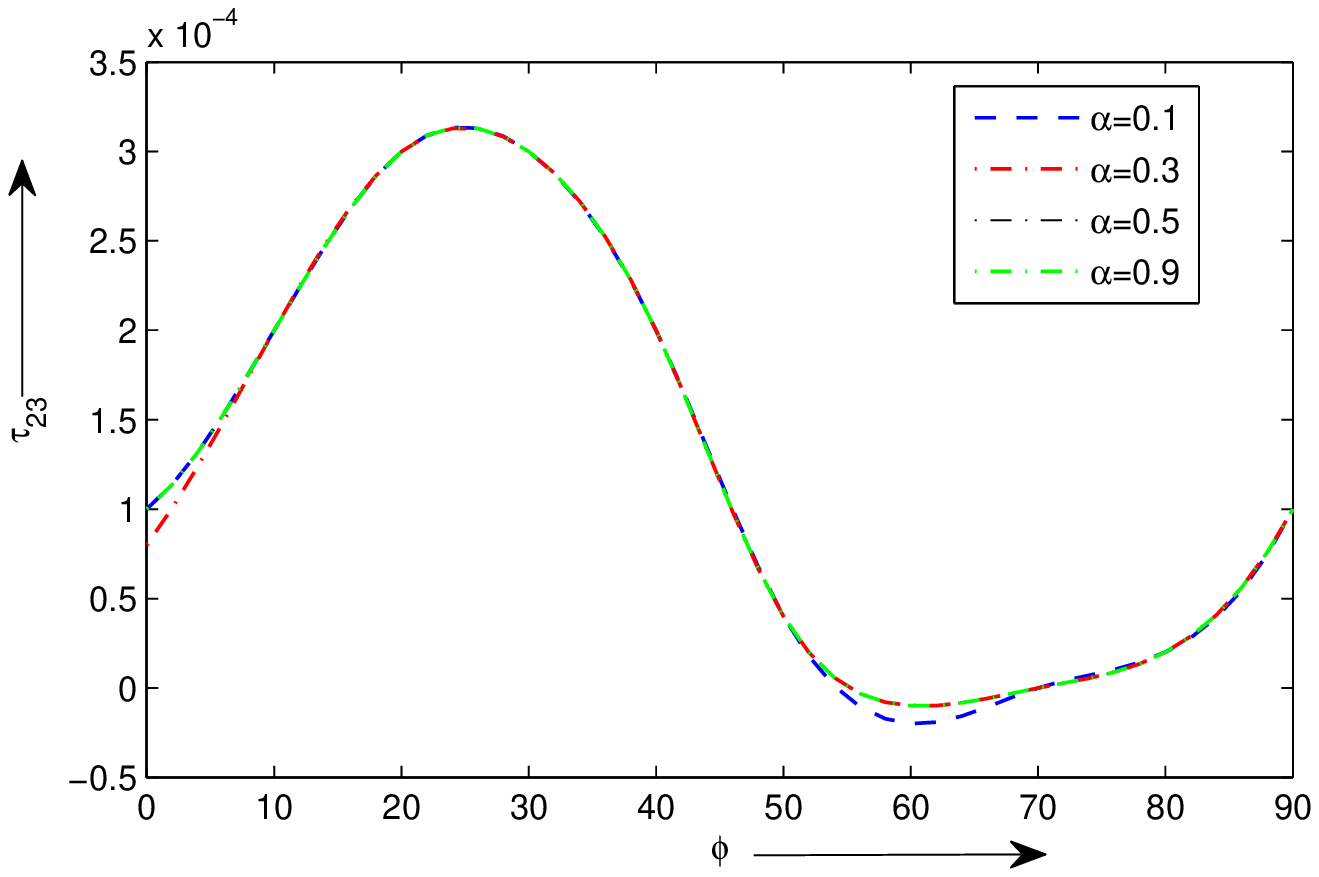}\\
Fig. 4~~ Distribution of ($\tau_{23}$) vs. $\phi$ \\

 \includegraphics[width=4.8in,height=3.5in ]{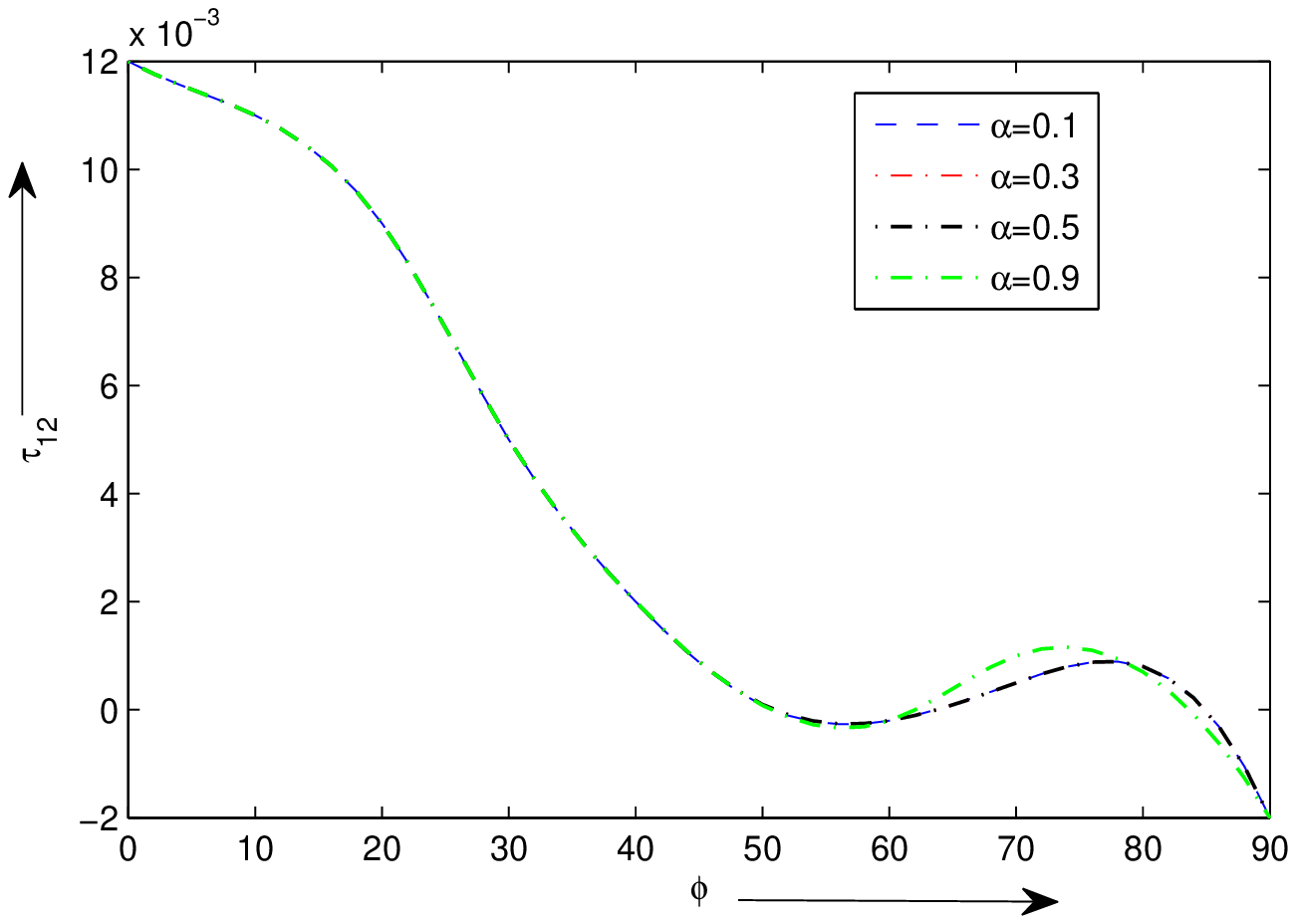}\\
Fig. 5 ~~ Distribution of ($\tau_{12}$) vs. $\phi$ \\
\end{center}

Figure 4 shows the behavior of stress  $\tau_{23}$ which is identical in nature for four values of $\alpha$. It attains the maximum values at $\phi=20^0$. Figure  5 depicts the distributions of stress $\tau_{12}$ versus the rotational angle $\phi$. The absolute value of $\tau_{12}$ decreases gradually as increase in $\phi$  in the region  $0^0 \le \phi \le 50^0$. and finally it vanishes at $\phi=90^0$.

   \begin{center}
\includegraphics[width=4.8in,height=3.5in ]{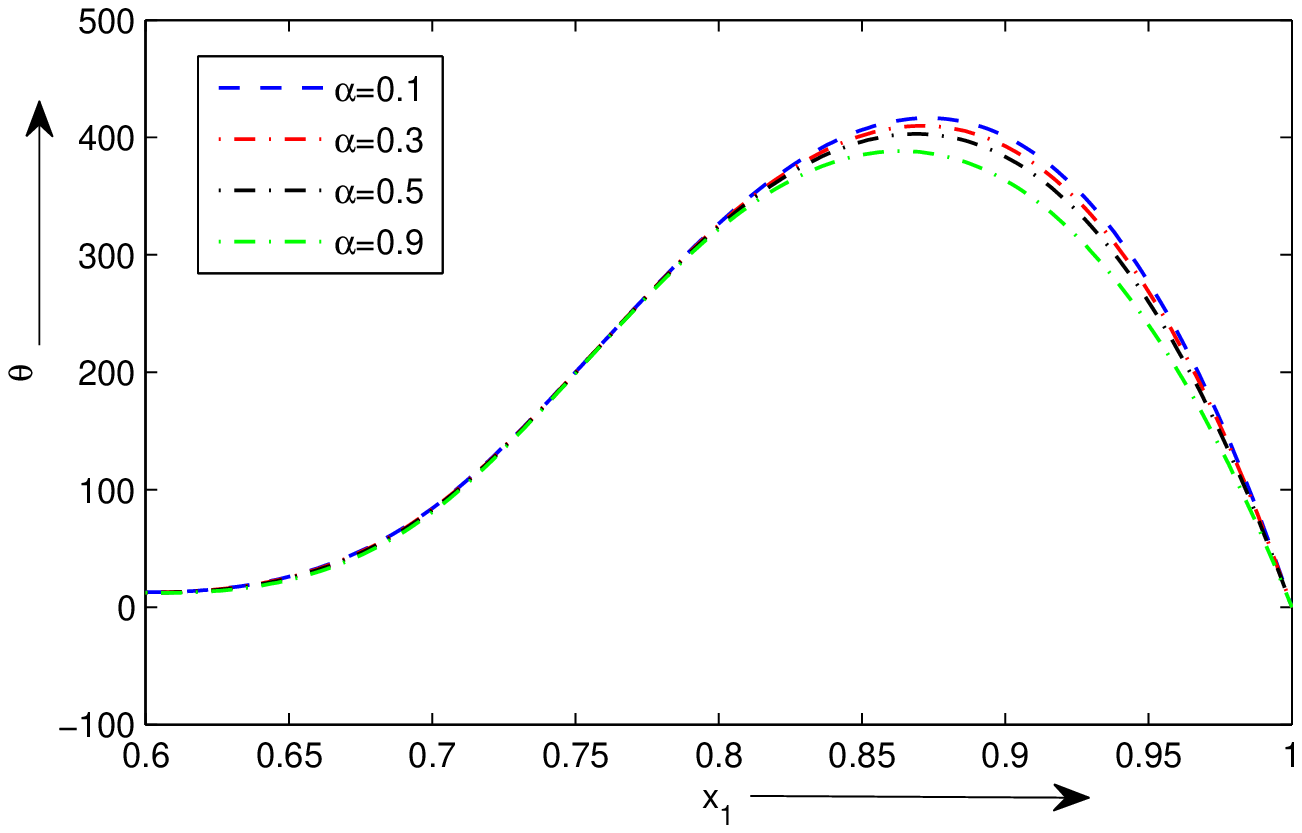}\\
Fig. 6~~Distribution of ($\theta$) vs. $x_1$ \\
\end{center}

\vspace*{4.5cm}
 \includegraphics[width=3.8in,height=2.5in ]{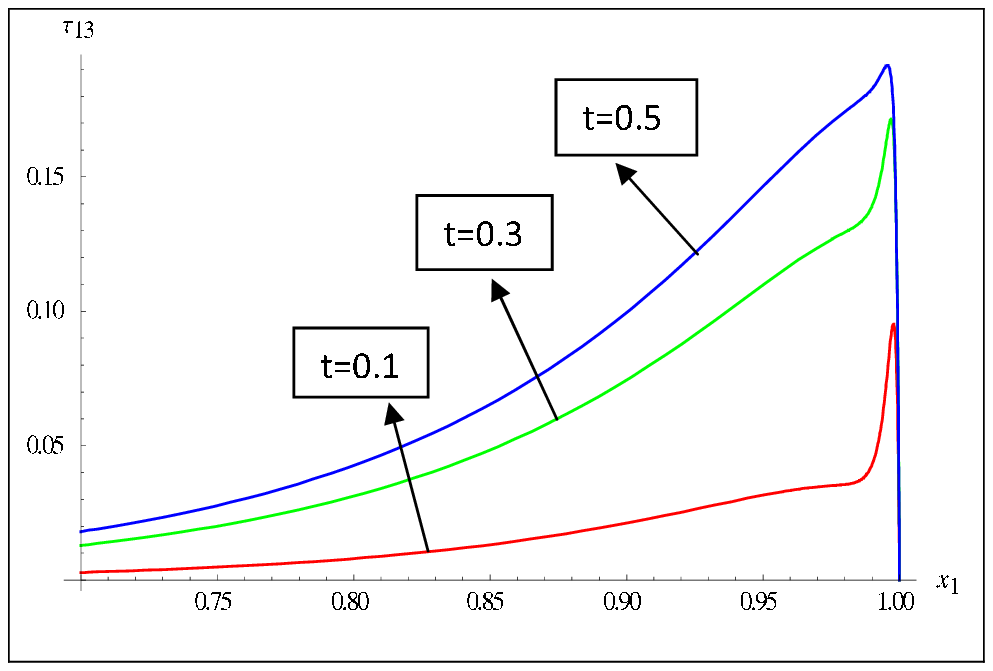}\\
Fig. 7~~Distribution of ($\tau_{13}$) vs. $x_1$\\

Figure 6 represents the variations of temperature versus space variable for fixed value of $\phi=80^0$ for different values of $\alpha$. It is observed that $\theta$ is maximum within the region $0\le x_1\le 0.9$ and finally it vanishes. Figure 7 represents the distribution of stress$\tau_{13}$ versus $x_1$ for fixed value of $\alpha=0.4$ and $\phi=10^0$. The characteristics of the curves remains same for all times. It is seen that for fixed values of $x_1$, stress component($\tau_{13}$) gradually increases as time increases. For fixed time($t$), it is also increased as increases of space variable($x_1$). From the figure, it is seen that stress($\tau_{13}$) is maximum near the plane surface of the plate and finally at the top of plate i.e., $x_1=1$, it vanishes.\\

\section{Conclusion}
A rigorous mathematical study of three-dimensional anisotropic thermoelastic analysis of simply supported rectangular plate with constant thickness and subjected to sinusoidal mechanical load on the top surface of the plate while other mechanical and thermal loads vanish on the top and bottom surface surface of the plate has been investigated using Green-Naghdi model-II. Displacement components, temperature distribution and stress components have been formulated analytically based on generalized three-dimensional thermoelastic theory. The non-linear heat conduction equation with Green-Naghdi's parameter has been analytically solved along with equation of motion and corresponding electro-magnetic field equations using normal mode analysis. The robust computations deals with-\\

      1.~~~~~~~The magnetic field, rotation and fractional order of heat equation plays an important role on all distributions of the field equations.\\

     2.~~~~~~~This problem can be solved for isotropic rectangular plate taking only two material constants ($\lambda$ and $\mu$, Lam$\grave{e}$ Constants) in spite of the material constants $c_{ij}$ in stress-strain relation and similarly, for $\alpha=1$, it may be solved for both the cases of isotropic and anisotropic rectangular plate with linear heat conduction equation.\\

     3.~~~~~~~The effect of the non-linear parameter for stresses and temperature are shown in Figures 3-6.\\

     4.~~~~~~~The analysis and results in this study should be more helpful in the branches of Geological Science and Geophysics considering the upper crust of Earth's surface as anisotropic elastic plate.\\


{\bf{References}}

\noindent{\bf{ Appendix\emph{}}}
\[a_{11} =\frac{1}{1+\varepsilon_1}[2ia\frac{c_{16} }{c_{11} } +2ib\frac{c_{15} }{c_{11} }] , a_{12} =\frac{1}{1+\varepsilon_1}[-a^{2} \frac{c_{66} }{c_{11} } -b^{2} \frac{c_{55} }{c_{11} } -2ab\frac{c_{56} }{c_{11} } -\varepsilon_2\omega ^{2}-\omega^2+\Omega_2^2] ,\]
 \[a_{21} =\frac{c_{16} }{(1+\varepsilon_1)c_{11} } ,
a_{22} =\frac{1}{1+\varepsilon_1}[ia\frac{c_{12} +c_{66} }{c_{11} } +ib\frac{c_{14} +c_{56} }{c_{11} }+ia\varepsilon_1] ,\]
\[a_{23} =\frac{1}{1+\varepsilon_1}[-a^{2} \frac{c_{26} }{c_{11} } -b^{2} \frac{c_{45} }{c_{11} } -ab\frac{c_{46} +c_{25} }{c_{11} }-\Omega_1\Omega_2] , a_{31} =\frac{c_{15} }{(1+\varepsilon_1)c_{11} } ,\]
\[a_{32} =\frac{1}{1+\varepsilon_1}[ia\frac{c_{14} +c_{56} }{c_{11} } +ib\frac{c_{13} +c_{55} }{c_{11} } +ib\varepsilon_1],\] \[a_{33} =\frac{1}{1+\varepsilon_1}[-a^{2} \frac{c_{46} }{c_{11} } -b^{2} \frac{c_{35} }{c_{11} } -ab\frac{c_{36} +c_{45} }{c_{11} }-2\omega\Omega_2 ],\]
\[ a_{34}=\frac{1+\tau_1\omega}{1+\varepsilon_1} , b_{11} =\frac{c_{16} }{c_{66} } ,
b_{12} =ia\frac{c_{12} +c_{66} }{c_{66} } +ib\frac{c_{14} +c_{56} }{c_{66} } ,\]
 \[b_{13} =-a^{2} \frac{c_{26} }{c_{66} } -b^{2} \frac{c_{45} }{c_{66} } -ab\frac{c_{46} +c_{25} }{c_{66} }-\Omega_1\Omega_2 , b_{21} =2ia\frac{c_{26} }{c_{66} } +2ib\frac{c_{46} }{c_{66} } , \]
\[b_{22} =-a^{2} \frac{c_{22} }{c_{66} } -b^{2} \frac{c_{44} }{c_{66} } -2ab\frac{c_{24} }{c_{66} } +\frac{c_{11}(\Omega_1^2-\omega ^{2}) }{c_{66} }  ;  b_{31} =\frac{c_{56} }{c_{66} } , b_{32} =ia\frac{c_{46} +c_{25} }{c_{66} } +ib\frac{c_{36} +c_{45} }{c_{66} } ,\]
 \[  b_{33} =-a^{2} \frac{c_{24} }{c_{66} } -b^{2} \frac{c_{34} }{c_{66} } -ab\frac{c_{23} +c_{44} }{c_{66} }+\frac{2c_{11}\omega\Omega_1}{c_{66}} ,
   b_{34} =\frac{ia\beta _{2} c_{11}(1+\tau_1\omega) }{c_{66} } ,  m_{11} =\frac{c_{15} }{c_{55} } ,\]
    \[m_{12} =ia\frac{c_{56} +c_{14} }{c_{55} } +ib\frac{c_{55} +c_{13} }{c_{55} }+ib\varepsilon_1\frac{c_{11}}{c_{55}} ,
    m_{22} =ia\frac{c_{25} +c_{46} }{c_{55} } +ib\frac{c_{45} +c_{36} }{c_{55} } ,\]
\[  m_{13} =-a^{2} \frac{c_{46} }{c_{55} } -b^{2} \frac{c_{35} }{c_{55} } -ab\frac{c_{45} +c_{36} }{c_{55} }+2\Omega_2\omega\frac{c_{11}}{c_{55}} , m_{21} =\frac{c_{56} }{c_{55} } , \]
  \[m_{23} =-a^{2} \frac{c_{24} }{c_{55} } -b^{2} \frac{c_{34} }{c_{55} } -ab\frac{c_{44} +c_{23} }{c_{55} }-\frac{c_{11}(\varepsilon_1 ab+2\Omega_1 \omega)}{c_{55}} ,m_{31} =2ia\frac{c_{45} }{c_{55} } +2ib\frac{c_{35} }{c_{55} } ,\]
   \[m_{32} =-a^{2} \frac{c_{44} }{c_{55} } -b^{2} \frac{c_{33} }{c_{55} } -2ab\frac{c_{34} }{c_{55} } -\frac{c_{11} }{c_{55} }(\Omega_1^2+\Omega_2^2- \omega ^{2}) , m_{33} =\frac{(1+\tau_1\omega)ib\beta _{3} c_{11} }{c_{55} } ,\]
 $h_{k1} =\frac{c_{k1} }{c_{11} };k=2(1)6, h_{k6} =\frac{i(bc_{k3} +ac_{k4} )}{c_{11} }, h_{k2} =\frac{c_{k6} }{c_{11} }, h_{k3} =\frac{c_{k5} }{c_{11} }, h_{k4} =\frac{i(bc_{k5} +ac_{k6} )}{c_{11} }, h_{k5} =\frac{i(bc_{k4} +ac_{k2} )}{c_{11} };k=1(1)6,$ $d_{1k} =a_{1k}+\frac{a_{21} (b_{31} m_{1k} -b_{1k} )}{1-b_{31} m_{21} } +\frac{a_{31} (m_{21} b_{1k} -m_{1k} )}{1-b_{31} m_{21} } ,k=1(1)8, where, a_{13}=a_{12}, a_{14}=a_{22}, a_{15}=a_{23}, a_{16}=a_{32}, a_{17}=a_{33}, a_{18}=0, $ $d_{2k} =(-b_{1k} -\frac{b_{31} (m_{21} b_{1k} -m_{1k} )}{1-b_{31} m_{21} } )-\frac{d_{1k} }{d_{11} } (-b_{11} -\frac{b_{31} (m_{21} b_{11} -m_{11} )}{1-b_{31} m_{21} } ),k=1(1)8, where,  b_{23}=b_{21}, b_{24}=b_{22}, b_{25}=b_{32}, b_{26}=b_{33}, b_{27}=0, b_{28}=b_{34},$
\[d_{31} =\frac{m_{11} d_{12} }{d_{11} } -m_{12} -m_{21} d_{21} , d_{32} =\frac{m_{11} d_{13} }{d_{11} } -m_{13} -m_{21} d_{22} ,\]
\[d_{33} =\frac{m_{11} d_{14} }{d_{11} } -m_{22} -m_{21} d_{23} , d_{34} =\frac{m_{11} d_{15} }{d_{11} } -m_{23} -m_{21} d_{24} , d_{35} =\frac{m_{11} d_{16} }{d_{11} } -m_{31} -m_{21} d_{25} ,\]
 \[d_{36} =\frac{m_{11} d_{17} }{d_{11} } -m_{32} -m_{21} d_{26} , d_{37} =-(\frac{m_{11} a _{34} }{d_{11} } +m_{21} d_{27} ), d_{38} =\frac{m_{11} d_{18} }{d_{11} } +m_{33} -m_{21} d_{28} , \]
\[\varepsilon _{1} =\omega ^{2-\alpha } \varepsilon {\rm (1+}\tau _{{\rm 2}} \omega ), \varepsilon _{2} =\omega ^{2-\alpha } \varepsilon {\rm (1+}\tau _{{\rm 2}} \omega )ia\beta _{2} , \varepsilon _{3} =\omega ^{2-\alpha } \varepsilon {\rm (1+}\tau _{{\rm 2}} \omega )ib\beta _{3}  ,\]
\[d_{41} =\omega^{2-\alpha}\varepsilon (1+\tau_2\omega) , d_{42} = d_{43} =0, d_{44} =\omega^{2-\alpha}\varepsilon \beta_2 (1+\tau_2\omega)ia , d_{45} =0,\]
\[ d_{46} =\omega^{2-\alpha}\varepsilon \beta_{3} (1+\tau_2\omega)ib , d_{47} =0, d_{48} =\omega ^{2-\alpha } (1+\tau _{2} \omega )+k_{2} a^{2} +k_{3} b^{2} , \]
\[g_{51} =-\frac{d_{13} }{d_{11} } , g_{52} =-\frac{d_{15} }{d_{11} } , g_{53} =-\frac{d_{17} }{d_{11} } , g_{54} =-\frac{d_{18} }{d_{11} } , g_{55} =-\frac{d_{12} }{d_{11} } , g_{56} =-\frac{d_{14} }{d_{11} } , g_{57} =-\frac{d_{16} }{d_{11} } ,\]
 $g_{58} =\frac{a_{34}}{d_{11} } ,    g_{61} =d_{22} , g_{62} =d_{24} , g_{63} =d_{26} , g_{64} =d_{28} , g_{65} =d_{21} , g_{66} =d_{23} , g_{67} =d_{25} , g_{68} =d_{27} ,$ $g_{71} =d_{32} , g_{72} =d_{34} , g_{73} =d_{36} , g_{74} =d_{38} , g_{75} =d_{31} , g_{76} =d_{33} , g_{77} =d_{35} , g_{78} =d_{37} , g_{81} =0,$ $g_{82} =d_{44} , g_{83} =d_{46} , g_{84} =d_{48} , g_{85} =d_{41} , g_{86} =g_{87} =g_{88} =0, f_{11} =g_{51} +\lambda g_{55} -\lambda ^{2} , f_{12} =g_{52} +\lambda g_{56} ,$ $f_{13} =g_{53} +\lambda g_{57} , f_{14} =g_{54} +\lambda g_{58} ,f_{21} =g_{61} +\lambda g_{65} ,
f_{22} =g_{62} +\lambda g_{66} -\lambda ^{2} , f_{23} =g_{63} +\lambda g_{67} ,$ $f_{24} =g_{64} +\lambda g_{68} ,f_{31} =g_{71} +\lambda g_{65} , f_{32} =g_{72} +\lambda g_{76} ,   f_{33} =g_{73} +\lambda g_{77} -\lambda ^{2} , f_{34} =g_{74} +\lambda g_{78} , f_{41} =\lambda g_{85} ,$ $f_{42} =g_{82} , f_{43} =g_{83} , f_{44} =g_{84} ,$ $L_{21}=\left[ g_{ij}\right]_{i=5(1)8,j=1(1)4}, L_{22}=\left[ g_{ij}\right]_{i=5(1)8,j=5(1)8}$;

  $\delta _{1} =(f_{24} f_{13} -f_{14} f_{23} )(f_{22} f_{33} -f_{32} f_{23} )-(f_{34} f_{23} -f_{24} f_{33} )(f_{12} f_{23} -f_{22} f_{13} )$,
$\delta _{2} =(f_{34} f_{23} -f_{24} f_{33} )(f_{11} f_{23} -f_{21} f_{13} )-(f_{24} f_{13} -f_{14} f_{23} )(f_{21} f_{33} -f_{31} f_{23} ),$
$\delta _{3} =(f_{12} f_{21} -f_{11} f_{22} )(f_{21} f_{34} -f_{31} f_{24} )-(f_{22} f_{31} -f_{21} f_{32} )(f_{11} f_{24} -f_{14} f_{21} ),$
$\delta _{4} =(f_{11} f_{23} -f_{21} f_{13} )(f_{22} f_{33} -f_{32} f_{23} )-(f_{12} f_{23} -f_{22} f_{13} )(f_{21} f_{33} -f_{31} f_{23} )$;\\
 $D_k=\left| R_{ij}\right|_{i,j=1(1)8}, R_{kj}=P_j,j\neq k,~and~D_9=\left| R_{ij}\right|_{i,j=1(1)8}~ where, ~R_{1j}=R_{5j}(0), R_{2j}=R_{5j}(h),
 R_{3j}=R_{6j}(0), R_{4j}=R_{6j}(h), R_{5j}=R_{1j}(0), R_{6j}=R_{1j}(h), R_{7j}=R_{7j}(0), R_{8j}=R_{7j}(h) ~for~ k=1(1)8$ ;
$A_j=D_j/D_9, j=1(1)8; P_j=0~for ~j=1(1)8~; ~j\neq 6 ~where~ P_6=- Sin[\pi x_2/c]Sin[\pi x_3/d]$;\\
 ${R_{kj} (x_{1} )=[(h_{k4} +h_{k1}\lambda _{j} )\{ (\delta _{1} )_{\lambda =\lambda _{j} }\} +(h_{k5} +h_{k2}\lambda _{j} )\{ (\delta _{2} )_{\lambda =\lambda _{j} } \}]e^{\lambda _{j} x_{1} }}$\\
  $+ [(h_{k6} +h_{k3}\lambda _{j})\{ (\delta _{3} )_{\lambda =\lambda _{j} } \} -\beta_k\{ (\delta _{4} )_{\lambda =\lambda _{j} } \} ]e^{\lambda _{j} x_{1} } ; \beta_1=1,\beta_4=\beta_5=\beta_6=0;$
${R_{7j} (x_1)= [(\delta _{4} )_{\lambda =\lambda _{j} } ] e^{\lambda _{j}x_{1} } ;}~for~ k=1(1)6, j=1(1)8.$ \\

\end{document}